\documentclass[aps,prb,preprint,showpacs,showkeys,floatfix]{revtex4}

\usepackage{graphicx,color}
\usepackage{multirow,slashbox}

\graphicspath{{figs/}}
\begin{document}

\title{Origin of Hinge-Like Mechanism in Single-Layer Black Phosphorus: the Angle-Angle Cross Interaction}

\author{Jin-Wu Jiang}
    \altaffiliation{Corresponding author: jiangjinwu@shu.edu.cn; jwjiang5918@hotmail.com}
    \affiliation{Shanghai Institute of Applied Mathematics and Mechanics, Shanghai Key Laboratory of Mechanics in Energy Engineering, Shanghai University, Shanghai 200072, People's Republic of China}

\date{\today}
\begin{abstract}

The single-layer black phosphorus is characteristic for its puckered configuration that possesses the hinge-like mechanism, which leads to the highly anisotropic in-plane Poisson's ratios and the negative out-of-plane Poisson's ratio. We reveal that the hinge-like mechanism can be attributed to the angle-angle cross interaction, which, combined with the bond stretching and angle bending interactions, is able to provide a good description of the mechanical properties in the single-layer black phosphorus. We also propose a nonlinear angle-angle cross interaction, which follows the form of the Stillinger-Weber potential and is advantageous for molecular dynamics simulations of single-layer black phosphorous under large deformations.

\end{abstract}
\maketitle
\pagebreak

\section{Introduction}

\begin{figure}[tb]
  \begin{center}
    \scalebox{1}[1]{\includegraphics[width=8cm]{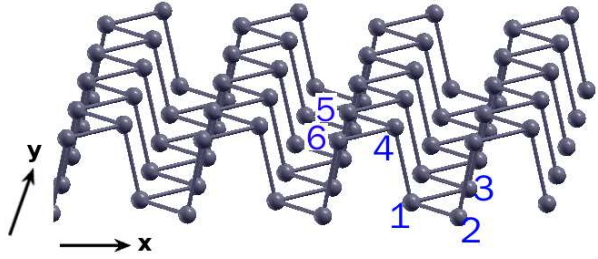}}
  \end{center}
  \caption{(Color online) Structure for SLBP. Atoms 1, 2, and 3 are in the bottom group, while atoms 4, 5, and 6 are in the top group.}
  \label{fig_cfg}
\end{figure}

The single-layer black phosphorus (SLBP) has anisotropic properties in the two in-plane directions due to its puckered atomic configuration shown in Fig.~\ref{fig_cfg}, in which the x and y axes are set in the directions perpendicular or parallel to the pucker. The Young's modulus in the y-direction is about four times larger than the Young's modulus in the x-direction.\cite{JiangJW2014bpyoung,QiaoJ2014nc,QinGarxiv14060261,WeiQ2014apl} The puckered configuration brings an interesting hinge-like mechanism for the SLBP; i.e., tension along the y-direction will generate a strong contraction in the x-direction. As a direct result of this hinge-like mechanism, the Poisson's ratio $\nu_{\rm xy}$ is much smaller than $\nu_{\rm yx}$.\cite{JiangJW2014bpnpr,WeiQ2014apl,ElahiM2014prb} Interestingly enough, the Poisson's ratio $\nu_{\rm yz}$ is negative owning to the hinge-like mechanism.\cite{JiangJW2014bpnpr,QinGarxiv14060261,ElahiM2014prb} We note that the Poisson's ratio $\nu_{\rm xy}=-\epsilon_y/\epsilon_x$ is corresponding to the tension of the SLBP along the x-direction, where $\epsilon_x$ and $\epsilon_y$ are the applied and resultant strains, respectively.

The hinge-like mechanism in SLBP has been discussed in terms of the structure deformation within first-principles calculations.\cite{JiangJW2014bpnpr,QinGarxiv14060261} Different from first-principles calculations, empirical potentials can provide intuitive explanations for physical or mechanical phenomena. In 1982, a valence force field (VFF) model was proposed to describe the interaction for SLBP under small linear deformations.\cite{KanetaC1982ssc} In a recent work, we suggested to simplify the VFF model by keeping major potential terms, and this simplified VFF model was used to derive parameters for the Stillinger-Weber (SW) potential.\cite{JiangJW2015sw}

However, there is still no explicit empirical potential term to describe the hinge-like mechanism in the SLBP. That is both in-plane Poisson's ratios ($\nu_{\rm xy}$ and $\nu_{\rm yx}$) calculated from these empirical potentials (including the VFF model, the simplified VFF model, and the SW potential) are much smaller than the value computed by the first-principles calculations.\cite{MidtvedtD2015} Furthermore, the negative out-of-plane Poisson's ratio obtained from first-principles calculations can not be reproduced by these three empirical potentials.

In this paper, we investigate the origin of the hinge-like mechanism in the SLBP, which results in the strong anisotropy of in-plane Poisson's ratios and the negativity of the out-of-plane Poisson's ratio. We find that the hinge-like mechanism can be captured by the angle-angle cross (AAC) interaction. The AAC interaction combined with the bond stretching and angle bending interactions, can provide a well description for mechanical properties of SLBP. Finally, we propose a nonlinear AAC potential following the form of the SW potential, which is advantageous for molecular dynamics simulations of SLBP under large deformations.

\section{Structure}
The structure for SLBP shown in Fig.~\ref{fig_cfg} has been identified by experiment.\cite{TakaoY1981physica} Phosphorous atoms are divided into the top group (including atoms 4, 5, and 6) and the bottom group (including atoms 1, 2, and 3). There are two bond lengths, i.e., the intra-group bond (eg. bond 1-2) $d_{1}=2.224$~{\AA} and the inter-group bond (eg. bond 1-4) $d_{2}=2.244$~{\AA}. These two bond lengths are very close to each other, so it can be assumed that both bonds have the same length of\cite{KanetaC1982ssc} $d=2.224$~{\AA}. The intra-group angle (eg. $\angle 213$) is $\theta_{213}^{0}=96.359^{\circ}$ and the inter-group angle (eg. $\angle 214$) is $\theta_{214}^{0}=102.09^{\circ}$.

\begin{table*}
\caption{Parameters of the VFF model. The second column shows the expression for each potential term. Parameters for the original VFF model\cite{KanetaC1982ssc} are shown in the third and fourth columns in different units. Parameters for the simplified VFF model are shown in the fifth column. Structural variables can be found in Fig.~\ref{fig_cfg}.}
\label{tab_vffm_slbp}
\begin{tabular*}{\textwidth}{@{\extracolsep{\fill}}|c|c|c|c|c|}
\hline 
\multirow{2}{*}{} & \multirow{2}{*}{expression} & \multicolumn{2}{c|}{original VFFM} & simplified VFFM\tabularnewline
\cline{3-5} 
 &  & force constant (dyne cm$^{-1}$) & force constant (eV$\AA^{-2}$) & force constant (eV$\AA^{-2}$)\tabularnewline
\hline 
\hline 
bond stretching & $\frac{K_{r}}{2}\left(\Delta r_{12}\right)^{2}$ & $0.1598\times10^{6}$ & 9.9715 & 9.9715\tabularnewline
\hline 
bond stretching & $\frac{K_{r}'}{2}\left(\Delta r_{14}\right)^{2}$ & $0.1516\times10^{6}$ & 9.4598 & 9.9715\tabularnewline
\hline 
angle bending & $\frac{K_{\theta}}{2}d_{1}^{2}\left(\Delta\theta_{213}\right)^{2}$ & $0.1725\times10^{5}$ & 1.0764 & 1.0764\tabularnewline
\hline 
angle bending & $\frac{K_{\theta}'}{2}d_{1}d_{2}\left(\Delta\theta_{214}\right)^{2}$ & $0.1497\times10^{5}$ & 0.9341 & 0.9341\tabularnewline
\hline 
bond bond cross & $K_{rr'}\left(\Delta r_{12}\right)\left(\Delta r_{13}\right)$ & $0.1772\times10^{5}$ & 1.1057 & 0\tabularnewline
\hline 
bond bond cross & $K_{rr'}'\left(\Delta r_{12}\right)\left(\Delta r_{14}\right)$ & $0.1772\times10^{5}$ & 1.1057 & 0\tabularnewline
\hline 
bond anlge cross & $K_{r\theta}d_{1}\left(\Delta r_{12}\right)\left(\Delta\theta_{213}\right)$ & $0.1155\times10^{5}$ & 0.7207 & 0\tabularnewline
\hline 
bond angle cross & $K_{r\theta}'\sqrt{d_{1}d_{2}}\left(\Delta r_{12}\right)\left(\Delta\theta_{214}\right)$ & $0.1155\times10^{5}$ & 0.7207 & 0\tabularnewline
\hline 
bond angle cross & $K_{r\theta}''\sqrt{d_{1}d_{2}}\left(\Delta r_{14}\right)\left(\Delta\theta_{214}\right)$ & $0.1155\times10^{5}$ & 0.7207 & 0\tabularnewline
\hline 
\end{tabular*}
\end{table*}

\section{VFF model}
In 1982, Kaneta et al. proposed a VFF model to describe the interaction between Phosphorous atoms for the SLBP in the linear deformation regime. There are two major terms in the VFF model for small bond variation $\Delta r$ and angle variation $\Delta\theta$,
\begin{eqnarray}
V_{r} & = & \frac{1}{2}K_{r}\left(\Delta r\right)^{2},\\
\label{eq_vffm1}
V_{\theta} & = & \frac{1}{2}K_{\theta}d_{1}d_{2}\left(\Delta\theta\right)^{2},
\label{eq_vffm2}
\end{eqnarray}
where $K_{r}$ and $K_{\theta}$ are force constant parameters. The $V_{r}$ term is the potential that captures a variation in the bond length $\Delta r$. The $V_{\theta}$ is for the potential corresponding to the variation of the angle $\Delta\theta$, where the angle $\theta$ is formed by two bonds of length $d_{1}$ and $d_{2}$. There are nine terms in this original model. Tab.~\ref{tab_vffm_slbp} shows the expressions and parameters for the VFF model.

The first four terms in the original VFF model govern the bond stretching and the angle bending motion styles in the SLBP. The last five terms in the original VFF model correspond to the cross interactions between bonds and angles. The bond stretching and angle bending are typical motion styles, so we suggest to simplify the original VFF model by keeping only the first four terms, while the last five terms are ignored. We make a further simplification by using the same force constant for the first two terms in the VFF model, considering that these two bonds $d_1$ and $d_2$ have almost the same length. Parameters for this simplified VFF model are shown in the last column in Tab.~\ref{tab_vffm_slbp}. We used this simplified VFF model in our previous work,\cite{JiangJW2015sw} in which all force constant parameters are rescaled by the same factor of 0.76. We do not perform such rescaling in the present work.

\begin{table}
\caption{Two-body (bond stretching) SW potential parameters for SLBP used by GULP. The expression is $V_{2}=Ae^{\left[\rho/\left(r-r_{max}\right)\right]}\left(B/r^{4}-1\right)$.}
\label{tab_sw2_gulp_slbp}
\begin{tabular}{|c|c|c|c|c|c|}
\hline 
 & $A$ (eV) & $\rho$ ($\AA$) & $B$ ($\AA^{4}$) & $r_{min}$ ($\AA$) & $r_{max}$ ($\AA$)\tabularnewline
\hline 
\hline 
P-P & 4.0266 & 0.5648 & 12.1100 & 0.0 & 2.79\tabularnewline
\hline 
\end{tabular}
\end{table}

\begin{table*}
\caption{Three-body (angle bending) SW potential parameters for SLBP used by GULP. The expression is $V_{3}=Ke^{\left[\rho_{1}/\left(r_{12}-r_{max12}\right)+\rho_{2}/\left(r_{13}-r_{max13}\right)\right]}\left(\cos\theta-\cos\theta_{0}\right)^{2}$. The first two lines are for intra-group angles. The last two lines are for inter-group angles.  P1 indicates atoms from the top group, while P2 represents atoms in the bottom group.}
\label{tab_sw3_gulp_slbp}
\begin{tabular*}{\textwidth}{@{\extracolsep{\fill}}|c|c|c|c|c|c|c|c|c|c|c|}
\hline 
 & $K$ (eV) & $\theta_{0}$ (degree) & $\rho_{1}$ ($\AA$) & $\rho_{2}$ ($\AA$) & $r_{min12}$ $(\AA)$ & $r_{max12}$ $(\AA)$ & $r_{min13}$ $(\AA)$ & $r_{max13}$ $(\AA)$ & $r_{min23}$ $(\AA)$ & $r_{max23}$ $(\AA)$\tabularnewline
\hline 
\hline 
P1-P1-P1 & 19.828 & 96.359 & 0.5648 & 0.5648 & 0.00 & 2.79 & 0.00 & 2.79 & 0.00 & 3.89\tabularnewline
\hline 
P2-P2-P2 & 19.828 & 96.359 & 0.5648 & 0.5648 & 0.00 & 2.79 & 0.00 & 2.79 & 0.00 & 3.89\tabularnewline
\hline 
P1-P1-P2 & 17.776 & 102.094 & 0.5648 & 0.5648 & 0.00 & 2.79 & 0.00 & 2.79 & 0.00 & 3.89\tabularnewline
\hline 
P2-P2-P1 & 17.776 & 102.094 & 0.5648 & 0.5648 & 0.00 & 2.79 & 0.00 & 2.79 & 0.00 & 3.89\tabularnewline
\hline 
\end{tabular*}
\end{table*}

\begin{table*}
\caption{SW potential parameters for SLBP used by LAMMPS. The two-body potential expression is $V_{2}=\epsilon A\left(B_L\sigma^{p}r_{ij}^{-p}-\sigma^{q}r_{ij}^{-q}\right)e^{\left[\sigma\left(r_{ij}-a\sigma\right)^{-1}\right]}$. The three-body potential expression is $V_{3}=\epsilon\lambda e^{\left[\gamma\sigma\left(r_{ij}-a\sigma\right)^{-1}+\gamma\sigma\left(r_{jk}-a\sigma\right)^{-1}\right]}\left(\cos\theta_{jik}-\cos\theta_{0}\right)^{2}$. The quantity tol in the last column is a controlling parameter in LAMMPS. P1 indicates atoms from the top group, while P2 represents atoms in the bottom group.}
\label{tab_sw_lammps_slbp}
\begin{tabular*}{\textwidth}{@{\extracolsep{\fill}}|c|c|c|c|c|c|c|c|c|c|c|c|}
\hline 
 & $\epsilon$ (eV) & $\sigma$ ($\AA$) & $a$ & $\lambda$ & $\gamma$ & $\cos\theta_{0}$ & $A$ & $B_{L}$ & $p$ & $q$ & $tol$\tabularnewline
\hline 
\hline 
P1-P1-P1 & 1.000 & 0.565 & 4.940 & 19.828 & 1.000 & -0.111 & 4.027 & 119.005 & 4 & 0 & 0.0\tabularnewline
\hline 
P2-P2-P2 & 1.000 & 0.565 & 4.940 & 19.828 & 1.000 & -0.111 & 4.027 & 119.005 & 4 & 0 & 0.0\tabularnewline
\hline 
P1-P2-P2 & 1.000 & 0.565 & 4.940 & 0.000 & 1.000 & -0.111 & 4.027 & 119.005 & 4 & 0 & 0.0\tabularnewline
\hline 
P2-P1-P1 & 1.000 & 0.565 & 4.940 & 0.000 & 1.000 & -0.111 & 4.027 & 119.005 & 4 & 0 & 0.0\tabularnewline
\hline 
P1-P1-P2 & 1.000 & 0.565 & 4.940 & 17.776 & 1.000 & -0.210 & 0.000 & 119.005 & 4 & 0 & 0.0\tabularnewline
\hline 
P1-P2-P1 & 1.000 & 0.565 & 4.940 & 17.776 & 1.000 & -0.210 & 0.000 & 119.005 & 4 & 0 & 0.0\tabularnewline
\hline 
P2-P2-P1 & 1.000 & 0.565 & 4.940 & 17.776 & 1.000 & -0.210 & 0.000 & 119.005 & 4 & 0 & 0.0\tabularnewline
\hline 
P2-P1-P2 & 1.000 & 0.565 & 4.940 & 17.776 & 1.000 & -0.210 & 0.000 & 119.005 & 4 & 0 & 0.0\tabularnewline
\hline 
\end{tabular*}
\end{table*}

\begin{figure}[tb]
  \begin{center}
    \scalebox{1}[1]{\includegraphics[width=8cm]{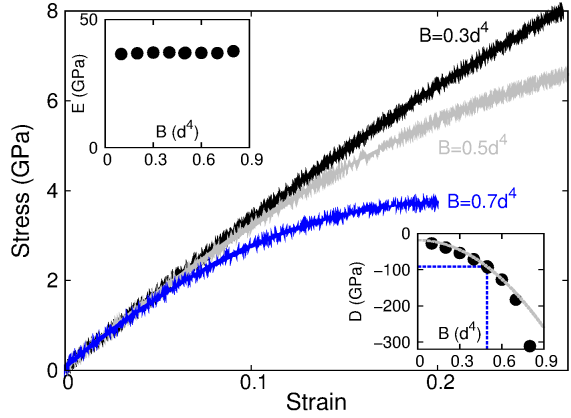}}
  \end{center}
  \caption{The effect of parameter B on the stress-strain relation for SLBP along the x direction at 1.0~K. The stress-strain curve is fitted to function $\sigma=E\epsilon+\frac{1}{2}D\epsilon^2$, with $E$ as the Young's modulus and $D$ as the TOEC. Left top inset shows that parameter $B$ has no effect on the elastic quantity, Young's modulus. However, the right bottom inset shows that the parameter $B$ has strong effect on the nonlinear property, TOEC, which is fitted to function $D=-18.2-298.8B^2$. The blue circle in the right bottom inset represents $D=-91.3$~{GPa} from the first-principles calculation,\cite{WeiQ2014apl} which helps to fix parameter $B=0.495d^4$ for the SW potential.}
  \label{fig_toec}
\end{figure}

\section{SW potential}
\label{sec_sw}

In a recent work, we proposed an analytic approach to parametrize the SW potential based on the VFF model for covalent materials.\cite{JiangJW2015sw} There are two-body and three-body interactions in the SW potential,
\begin{eqnarray}
V_{2} & = & Ae^{[\rho/\left(r-r_{max}\right)]}\left(B/r^{4}-1\right),
\label{eq_sw2}\\
V_{3} & = & Ke^{[\rho_{1}/\left(r_{12}-r_{max12}\right)+\rho_{2}/\left(r_{13}-r_{max13}\right)]}\left(\cos\theta-\cos\theta_{0}\right)^{2},\nonumber\\
\label{eq_sw3}
\end{eqnarray}
where $V_{2}$ corresponds to the bond stretching and $V_{3}$ associates with the angle bending. The cut-offs $r_{\rm max}$, $r_{\rm max12}$ and $r_{\rm max13}$ are geometrically determined by the material's structure. There are five unknown geometrical parameters, i.e., $\rho$ and $B$ in the two-body $V_2$ term and $\rho_1$, $\rho_2$, and $\theta_0$ in the three-body $V_3$ term, and two energy parameters $A$ and $K$. We obtained an analytic constraint for parameters in the SW potential,
\begin{eqnarray}
\rho & = & \frac{-4B\left(d-r_{max}\right)^{2}}{\left(Bd-d^{5}\right)}.
\label{eq_rho}
\end{eqnarray}
Following this parameterization procedure, we can utilize the simplified VFF model in the last column in Tab.~\ref{tab_vffm_slbp} to derive parameters for the SW potential. These SW parameters used by GULP\cite{gulp} are shown in Tabs.~\ref{tab_sw2_gulp_slbp} and ~\ref{tab_sw3_gulp_slbp}. SW potential parameters used by LAMMPS\cite{Lammps} are listed in Tab.~\ref{tab_sw_lammps_slbp}. The SW potential script for LAMMPS can be found in the appendix. 

The determination of $B$ is illustrated in Fig.~\ref{fig_toec}. The parameter $B$ has no effect on the elastic property, such as the Young's modulus, as shown by the left top inset in Fig.~\ref{fig_toec}. However, the parameter $B$ has strong effect on the third order elastic constant, which can be fitted to the function $D=-22.5-307.2B^2$. Using this relationship between the third order elastic constant and parameter $B$, we obtain the parameter $B=0.495d^4$ corresponding to $D=-91.3$~{GPa} from the first-principles calculations.\cite{WeiQ2014apl} We note that the relationship $B=0.495d^4$ in the present work is slightly different from $B=0.584d^4$ in our previous work,\cite{JiangJW2015sw} as the parameters of the simplified VFF model are slightly different in these two works.

To obtain these stress-strain relations, we use LAMMPS to perform molecular dynamics simulations for the tension of the SLBP of dimension $26.3\times 29.8$~{\AA} at 1.0~K along the x direction. Periodic boundary conditions are applied in both x and y directions. The structure is thermalized to the thermal steady state with the NPT (constant particle number, constant pressure, and constant temperature) ensemble for 100~ps by the Nos\'e-Hoover\cite{Nose,Hoover} approach. After thermalization, the SLBP is stretched in one direction at a strain rate of $10^8$~{s$^{-1}$}, and the stress in the lateral direction is allowed to be fully relaxed. We have used the inter-layer space of 5.24~$\AA$ as the thickness of the SLBP in the computation of the strain energy density.

\section{AAC interaction}
\label{sec_aac}

The puckered configuration of SLBP results in strong anisotropy in most mechanical properties. For example, the in-plane Poisson's ratio $\nu_{yx}$ is much larger than the other in-plane Poisson's ratio $\nu_{xy}$. More specifically, first-principles calculations predicted $\nu_{yx}=0.62$ and $\nu_{xy}=0.17$ in Ref~\onlinecite{WeiQ2014apl}, or $\nu_{yx}=0.81$ and $\nu_{xy}=0.24$ in Ref~\onlinecite{ElahiM2014prb}, or $\nu_{yx}=0.75$ and $\nu_{xy}=0.16$ in Ref~\onlinecite{JiW2015pc}. As another example, the Young's modulus in the x-direction is much smaller than that in the y-direction. The Young's modulus in the x and y-directions from first-principles calculations are 28.9~{Nm$^{-1}$} and 101.6~{Nm$^{-1}$} in Ref~\onlinecite{QiaoJ2014nc}, or 24.3~{Nm$^{-1}$} and 80.2~{Nm$^{-1}$} in Ref~\onlinecite{QinGarxiv14060261}, or 24.4~{Nm$^{-1}$} and 92.1~{Nm$^{-1}$} in Ref~\onlinecite{WeiQ2014apl}.

The puckered configuration possesses the hinge-like mechanism, which leads to a negative value for the out-of-plane Poisson's ratio $\nu_{yz}$. For example, $\nu_{yz}=-0.027$ from Ref~\onlinecite{JiangJW2014bpnpr}, or $\nu_{yz}=-0.059$ from Ref~\onlinecite{QinGarxiv14060261}, or $\nu_{yz}=-0.09$ from Ref~\onlinecite{ElahiM2014prb}. Actually, the negativity of the out-of-plane Poisson's ratio is closely related to the strong anisotropy of the in-plane Poisson's ratios.

The Poisson's ratio values calculated from the SW potential in Sec.~\ref{sec_sw} are much smaller than the above first-principles calculations. It is because the Poisson's ratios from the original VFF model ($\nu_{yx}=0.012$ and $\nu_{xy}=0.054$) are smaller than the first-principles calculations.\cite{MidtvedtD2015} Smaller Poisson's ratios are thus expected for the SW potential, which is derived based on the VFF model. Similarly, the out-of-plane Poisson's ratio from the SW potential in Sec.~\ref{sec_sw} is positive instead of negative as predicted by first-principles calculations. It implies that the hinge-like mechanism has not been captured by the SW potential or the VFF model yet.

\begin{figure}[tb]
  \begin{center}
    \scalebox{1}[1]{\includegraphics[width=8cm]{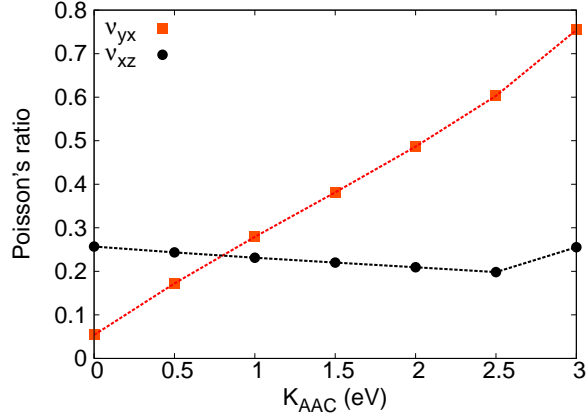}}
  \end{center}
  \caption{(Color online) The Poisson's ratio $\nu_{yx}$ and $\nu_{xz}$ for SLBP described by the SW potential combined with the AAC interaction with different parameters $K_{\rm aac}$.}
  \label{fig_poisson1}
\end{figure}

\begin{figure}[tb]
  \begin{center}
    \scalebox{1}[1]{\includegraphics[width=8cm]{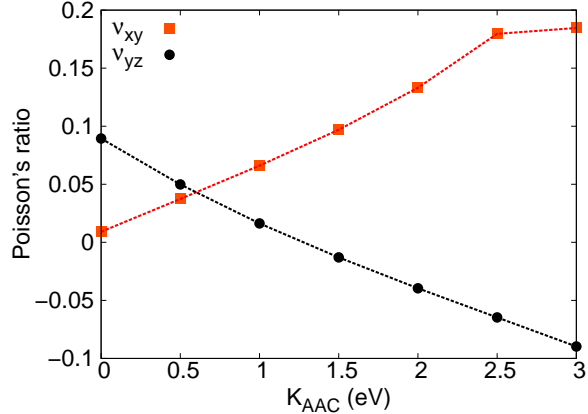}}
  \end{center}
  \caption{(Color online) The Poisson's ratio $\nu_{xy}$ and $\nu_{yz}$ for SLBP described by the SW potential combined with the AAC interaction with different parameters $K_{\rm aac}$.. The Poisson's ratio $\nu_{yz}$ becomes negative for $K_{\rm aac}>1.5$~{eV}.}
  \label{fig_poisson2}
\end{figure}

After carefully examining the puckered structure of the SLBP shown in Fig.~\ref{fig_cfg}, we suggest the following AAC interaction to describe the hinge-like mechanism for the SLBP,
\begin{eqnarray}
V_{\rm aac} & = & K_{\rm aac}\left(\theta_{213}-\theta_{213}^{0}\right)\left(\theta_{214}-\theta_{214}^{0}\right),
\label{eq_aac}
\end{eqnarray}
where $K_{\rm aac}$ is the force constant for this interaction. It is obvious that this AAC interaction is missed in the original VFF model shown in Tab.~\ref{tab_vffm_slbp}. A direct effect of the AAC interaction is to couple the variations of the intra-group angles like $\theta_{213} $ and the inter-group angles like $\theta_{214}$ in Fig.~\ref{fig_cfg}. The intra-group angle $\theta_{213}$ increases during the tensile stretching of the SLBP in the y-direction. The function of the AAC interaction is thus to decrease the inter-group angle $\theta_{214}$, resulting in a strong contraction in the x-direction. That is the value of the in-plane Poisson's ratio $\nu_{yx}$ will be greatly increased by adding the AAC interaction. This effect for the AAC interaction actually actuates the hinge-like mechanism of the puckered structure in the SLBP. Due to this hinge-like mechanism, the out-of-plane Poisson's ratio $\nu_{yz}$ can turn to negative after the AAC interaction is strong enough.

The AAC interaction has been included in the investigation of mechanical properties for the polyethylene crystal,\cite{KarasawaN1991jpc,KarasawaN1992mm} the silicon nitride ceramics,\cite{WendelJA1992jcp} and the urea and melamine.\cite{MeierRJ1995jpc} The AAC interaction was also used in the CFF91 force field for the numerical simulations of the prion protein fragment.\cite{MaB2002pnas,KuwataK2003pnas}

We thus suggest to describe the interaction of the SLBP by the SW potential presented in Sec.~\ref{sec_sw} combined with the AAC interaction. Figs.~\ref{fig_poisson1} and ~\ref{fig_poisson2} verify the effect of the AAC interaction on the Poisson's ratios in the SLBP. Fig.~\ref{fig_poisson1} shows that the in-plane Poisson's ratio $\nu_{yx}$ increases almost linearly with increasing the strength of the AAC interaction, while the Poisson's ratio $\nu_{xz}$ is almost not affected. Fig.~\ref{fig_poisson2} shows that the other in-plane Poisson's ratio $\nu_{xy}$ also increases linearly with increasing the strength of the AAC interaction. It is quite interesting that the out-of-plane Poisson's ratio $\nu_{yz}$ decreases with increasing the strength of the AAC interaction, and $\nu_{yz}$ becomes negative for $K_{\rm aac}>1.5$~{eV}.

The Poisson's ratio and the Young's modulus presented in this section are computed using the GULP package,\cite{gulp} in which the AAC interaction in Eq.~(\ref{eq_aac}) has been implemented. The AAC interaction in Eq.~(\ref{eq_aac}) has not been implemented in LAMMPS.\cite{Lammps} The SLBP is stretched by an uniaxial strain, and the stretched system is relaxed by the conjugate gradient energy minimization approach. During the minimization procedure, the size of the SLBP in the uniaxial strain direction is kept unchanged while all other degrees of freedom are allowed to be relaxed. The Poisson's ratios are calculated using the relationship between the applied longitudinal strain and the resultant lateral strain. The Young's modulus is obtained from the strain dependence of the strain energy.

\begin{table}
\caption{The effects of the AAC interaction on mechanical properties in the SLBP. The interaction of the SLBP is described by the SW potential combined with the AAC interaction.}
\label{tab_aac_effect}
\begin{tabular}{|c|c|c|c|c|c|c|}
\hline 
$K_{aac}$ (eV) & $E_{x}$ ($Nm^{-1}$) & $E_{y}$ ($Nm^{-1}$) & $\nu_{xy}$ & $\nu_{yx}$ & $\nu_{xz}$ & $\nu_{yz}$\tabularnewline
\hline 
\hline 
0.0 & 16.8 & 75.1 & 0.009 & 0.054 & 0.257 & 0.089\tabularnewline
\hline 
2.0 & 19.0 & 66.7 & 0.133 & 0.486 & 0.209 & -0.040\tabularnewline
\hline 
\end{tabular}
\end{table}

\begin{figure}[tb]
  \begin{center}
    \scalebox{1}[1]{\includegraphics[width=8cm]{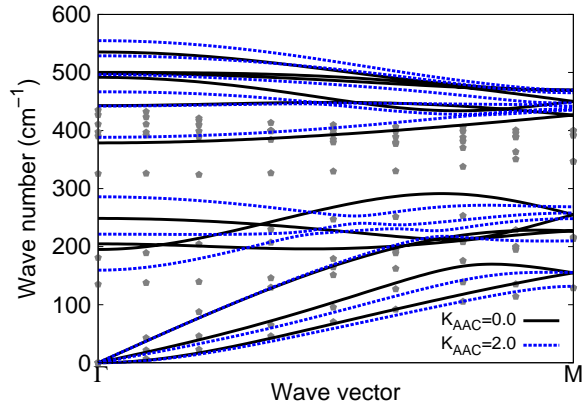}}
  \end{center}
  \caption{(Color online) Phonon spectrum for SLBP along $\Gamma$M from the SW potential combined with the AAC interaction with $K_{\rm AAC}=0.0$ and 2.0~{eV}. Phonon frequencies from the {\it ab initio} calculation\cite{ZhuZ2014prl} are shown as pentagons.}
  \label{fig_phonon}
\end{figure}

After evaluating the AAC effect on all quantities, we suggest to adopt $K_{\rm aac}=2.0$~{eV} as the force constant value for the the AAC interaction. The Young's modulus and the Poisson's ratio are shown in Tab.~\ref{tab_aac_effect} for $K_{\rm aac}=0$~{eV} and 2.0~{eV}. Using $K_{\rm aac}=2.0$~{eV}, the Young's modulus is 19.0~{Nm$^{-1}$} and 66.7~{Nm$^{-1}$} along the x and y-directions, respectively. These values are comparable with the above first-principles calculations. All of these four Poisson's ratios are comparable with the above first-principles calculations. In particular, the Poisson's ratio $\nu_{yz}$ in the out-of-plane direction is negative, and its value falls in the range for $\nu_{\rm yz}$ from the first-principles calculations. Fig.~\ref{fig_phonon} shows the phonon dispersion for SLBP, in which the interaction is described by the SW potential combined with the AAC interaction. The phonon dispersion for $K_{\rm aac}=2.0$~{eV} is in good agreement with the first-principles calculations in the low-frequency regime. The optical branches in the high-frequency regime from the SW combined with the AAC interaction are slightly higher than the first-principles calculations.

\section{AAC-SW interaction}
In Sec.~\ref{sec_aac}, we have suggested to describe the interaction by the SW potential combined with the AAC interaction, in which the AAC interaction captures the hinge-like mechanism for the SLBP. The AAC interaction shown in Eq.~(\ref{eq_aac}) is a linear potential, so it can not be used for the simulation of the SLBP under large deformations. We thus suggest the following nonlinear AAC interaction for the molecular dynamics simulations of SLBP under large deformations,
\begin{eqnarray}
V_{aac-sw} & = & K_{aac-sw}e^{[\frac{\rho_{1}}{r_{12}-r_{max12}}+\frac{\rho_{2}}{r_{13}-r_{max13}}+\frac{\rho_{3}}{r_{14}-r_{max14}}]}\nonumber\\
 &  & \left(\cos\theta_{213}-\cos\theta_{213}^{0}\right)\left(\cos\theta_{214}-\cos\theta{}_{214}^{0}\right).
\label{eq_aac-sw}
\end{eqnarray}
We will refer to this nonlinear AAC interaction as the AAC-SW interaction, as it follows the same mathematical format of the SW potential. That is this is essentially an angle-angle cross interaction following the same form of the SW potential. The force constant parameter $K_{\rm aac-sw}$ can be determined by the linear AAC interaction in Eq.~(\ref{eq_aac}). By equating Eq.~(\ref{eq_aac-sw}) to Eq.~(\ref{eq_aac}) at the equilibrium structure, we obtain
\begin{eqnarray}
K_{aac-sw} & = & \frac{K_{aac}}{e^{[3\rho/\left(d-r_{max12}\right)]}\sin\theta_{213}^{0}\sin\theta_{214}^{0}}\nonumber\\
& = & 40.8366~{\rm eV}.
\label{eq_Kaac-sw}
\end{eqnarray}
The AAC-SW potential proposed here is advantageous for molecular dynamical simulations under large deformations, as it naturally contains the nonliner component. The AAC-SW potential has the same effect as the AAC interaction for small linear deformations. Similar as the linear AAC interaction, the AAC-SW potential enables the hinge-like mechanism of the SLBP. It should be noted that the AAC-SW potential is also applicable to describe the hinge-like mechanism for other materials with similar puckered configuration like SnSe. The AAC-SW has not been implemented in either GULP or LAMMPS simulation packages yet. We are currently contacting developers of both packages for possible implementations.

\section{conclusion}
In conclusion, we discuss the relation between the AAC interaction and the hinge-like mechanism in SLBP, which is the origin for the negative Poisson's ratio in the out-of-plane direction. The AAC interaction combined with the VFF model can provide an overall well description for mechanical properties of the SLBP. In particular, these two in-plane Poisson's ratios are highly anisotropic and the out-of-plane Poisson's ratio is negative, which agrees with the first-principles predictions. We propose the nonlinear AAC-SW potential for molecular dynamics simulations of the SLBP under large deformations. The nonlinear AAC-SW potential will be useful in characterizing the hinge-like mechanism of SLBP and other materials with similar puckered configuration.

\textbf{Acknowledgements} The work is supported by the Recruitment Program of Global Youth Experts of China, the National Natural Science Foundation of China (NSFC) under Grant No. 11504225 and the start-up funding from Shanghai University.

\appendix

\section{Stillinger-Weber potential script for LAMMPS}

\# This is the sw.bp file for LAMMPS.

\# these entries are in LAMMPS metal units:

\# epsilon = eV; sigma = Angstroms

\# other quantities are unitless

\# format of a single entry (one or more lines):

\# element 1, element 2, element 3, 

\# epsilon, sigma, a, lambda, gamma, costheta0, A, B, p, q, tol

\# intra-group SW2 and SW3

T T T    1.000   0.565   4.940      19.828   1.000  -0.111   4.027 119.005  4  0 0.0

B B B    1.000   0.565   4.940      19.828   1.000  -0.111   4.027 119.005  4  0 0.0

\# inter-group SW2

T B B    1.000   0.565   4.940       0.000   1.000  -0.111   4.027 119.005  4  0 0.0

B T T    1.000   0.565   4.940       0.000   1.000  -0.111   4.027 119.005  4  0 0.0

\# inter-group SW3

T T B    1.000   0.565   4.940      17.776   1.000  -0.210   0.000 119.005  4  0 0.0

T B T    1.000   0.565   4.940      17.776   1.000  -0.210   0.000 119.005  4  0 0.0

\# inter-group SW3

B B T    1.000   0.565   4.940      17.776   1.000  -0.210   0.000 119.005  4  0 0.0

B T B    1.000   0.565   4.940      17.776   1.000  -0.210   0.000 119.005  4  0 0.0

%

\end{document}